\documentclass[epj]{webofc}
\usepackage[varg]{txfonts}   % Web of Conferences font
\usepackage{graphicx,color} % for figures
\usepackage{bm}       % bold math 
\usepackage{amsmath}  
\usepackage{amssymb}
\usepackage[utf8]{inputenc}
\woctitle{21st International Conference on Few-Body Problems in Physics}
\begin{document}
\title{The Falsification of Nuclear Forces}
\author{R. Navarro Perez\inst{1}\fnsep\thanks{\email{navarroperez1@llnl.gov}} \and
        J. E. Amaro\inst{2}\fnsep\thanks{\email{amaro@ugr.es}} \and
        E. Ruiz Arriola\inst{2}\fnsep\thanks{\email{earriola@ugr.es}}
}

\institute{Nuclear and Chemical Science Division, Lawrence Livermore
  National Laboratory\\ Livermore, California 94551, USA
\and
Departamento de F\'{\i}sica At\'omica, Molecular y Nuclear
  and \\ Instituto Carlos I de F{\'\i}sica Te\'orica y Computacional,
  Universidad de Granada \\ E-18071 Granada, Spain
          }

\abstract{ We review our work on the statistical uncertainty analysis
  of the NN force. This is based on the Granada-2013 database  where a
  statistically meaningful partial wave analysis comprising a total of
  6713 np and pp published scattering data from 1950 till 2013 below
  pion production threshold has been made. We stress the necessary
  conditions required for a correct and self-consistent statistical
  interpretation of the discrepancies between theory and experiment
  which enable a subsequent statistical error propagation and
  correlation analysis }
\maketitle
\section{Introduction}
\label{intro}

Error propagation and uncertainty quantification have recently become
a central topic in nuclear physics~\cite{Ireland:0954,
  Dobaczewski:2014jga, McDonnell:2015sja, Furnstahl:2014xsa,
  Carlsson:2015vda, Perez:2014bua}. In the particular field of
phenomenological Nucleon-Nucleon (NN) interactions uncertainties can
be classified as statistical or systematic\footnote{Numerical
  uncertainties are also present but can be made small enough to be
  dominated by the other two.}. Statistical uncertainties are the
result of unavoidable random fluctuations during the experimental
process. Most NN scattering measurements consist of counting events
which corresponds to a Poisson distribution. If the number of events
is large enough then the distribution can be safely approximated as a
normal one. This allows to fix the parameters of a phenomenological
potential via the usual chi square minimization process to reproduce
the collection of NN scattering data. As a consequence the statistical
uncertainty of the experimental data propagates into the fitting
parameters in the form of a confidence region in which the parameters
are allowed to vary and still give an accurate description of the
data. Such confidence region can be easily determined by the
parameters' covariance matrix if the assumption of normally
distributed residuals is fulfilled.  Systematic uncertainties are a
consequence of our lack of knowledge of the actual form of the NN
potential and the assumptions that have to be made in order to give a
representation of the NN interaction. Some potentials are separable in
momentum space while others are not, some are energy dependent and
others range from fully local to different types non-localities in
coordinate space. Even though most of the potentials are fitted to the
same type of experimental NN scattering data, their predictions of
unmeasured scattering observables or nuclear structure properties can
sometimes be incompatible. This residual incompatibilities beyond
statistical consistency and equivalence is what we refer as systematic
uncertainties. In this contribution we review our determination of the
NN interaction statistical uncertainties along with the necessary
conditions for such type of analysis.

\section{Coarse graining and the delta-shell potential}
\label{sec:CoarseGraining}
Coarse graining embodies the Wilsonian
renormalization~\cite{Wilson:1973jj} concept and represents a very
reliable tool to simplify the description of pp and np scattering data
while still retaining all the relevant information of the interaction
up to a certain energy range set by the de Broglie wavelength of the
most energetic particle considered. The $V_{\mathrm{low} k}$
potentials in momentum space are a good example of an implementation
of coarse graining by removing the high-momentum part of the
interaction~\cite{Bogner:2003wn,Bogner:2009bt}.  In 1973 Aviles
introduced the delta-shell (DS) potential in the context of NN
interactions~\cite{Aviles:1973ee}. In~\cite{NavarroPerez:2011fm} the
DS potential was used to implement coarse graining in coordinate space
via a local potential that samples the np interaction at certain
concentration radii $r_i$ by
\begin{equation}
V(r) = \sum_{i=1}^n \frac{\lambda_i}{2\mu}\delta(r-r_i),
\end{equation}
where $\mu$ is the reduced mass and $\lambda_i$ are strength
coefficients. After fitting the $\lambda_i$ parameters to np
phase-shifts the properties and form factors of the deuteron were
calculated. A variational method with harmonic oscillator
wave functions was used to calculate upper bounds to the binding
energy of the double-closed shell nuclei $^4$He, $^{16}$O and
$^{40}$Ca~\cite{NavarroPerez:2011fm}.
%% Effective interactions were
%% analyzed in light of the DS potential as a function of the maximum
%% fitting laboratory frame energy $T_{\mathrm{LAB}}$
%% in~\cite{NavarroPerez:2013iwa}. 

\section{Description of NN scattering data}
In order to quantify the statistical uncertainties of the NN
interaction a fit to experimental data becomes necessary. The usual
first step to fit a phenomenological potential to reproduce scattering
phase-shifts is not sufficient to get an accurate description of the
actual experimental data. As was recently shown
in~\cite{Perez:2014bua}, the local chiral effective potential
of~\cite{Gezerlis:2014zia} fitted to phase-shifts yields a
significantly large $\chi^2/{\rm d.o.f.}$ value when compared to
experimental scattering data. However it is possible that small
readjustments of the potential parameters have a significant impact in
lowering the total $\chi^2$. Given the wide applicability of this type
of interactions~\cite{Lynn:2014zia, Gandolfi:2014ewa} a full fledged
fit to NN experimental scattering data would be of great interest.

Historically, a successful description of the complete database with
$\chi^2/{\rm d.o.f.}  \lesssim 1$ has never been possible. The
potentials and PWA were gradually improved over time by explicitly
including different physical effects like OPE in the long range part,
charge symmetry breaking in the central channel and electromagnetic
interactions among others. The first PWA with $\chi^2/{\rm
  d.o.f.}\lesssim 1$ was obtained in 1993 when the Nijmegen group
introduced the $3\sigma$ criterion to exclude over $1000$ inconsistent
data\cite{Stoks:1993tb}. The $3\sigma$ criterion deals with possible
over and underestimations of the statistical uncertainties by
excluding data sets with improbably high or improbably low values of
$\chi^2$ (for a clear description of this process
see~\cite{Gross:2008ps}). However, this method identifies only
inconsistencies between individual data sets and a model trying to
describe the complete database. To improve on this method, so that
inconsistencies between each data set and the rest of the database can
be found, the $3\sigma$ criterion was applied iteratively to the
\emph{complete} database and the potential parameters were refitted to
the \emph{accepted} data sets until no more data are excluded or
recovered~\cite{Perez:2013jpa}. The self-consistent data base obtained
with this procedure contains $6713$ experimental points and recovers
$300$ of initially discarded data with the usual $3\sigma$ criterion
\cite{GranadaDB}.  Although the $300$ extra data do not significantly
change the potential parameters, their inclusion can only improve the
estimate of statistical errors. A simultaneous fit to pp and np
scattering data was made representing the short range part of the
interaction with a DS potential and OPE for the long range part. The
fit requires a total of $46$ parameters and yields $\chi^2/{\rm
  d.o.f.} = 1.04$ to the self-consistent data
base~\cite{Perez:2013mwa,Perez:2013jpa}.

\section{Description of NN scattering errors}
On a more fundamental level, any chi-squared distribution can be used
to test goodness of fit, provided that the experimental data can be
assumed to have a normal distribution~\cite{evans2009probability,
  james2006statistical}. If the residuals defined as
\begin{equation}
 R_i = \frac{O_i^{\rm exp}-O_i^{\rm theor}}{\Delta O_i^{\rm exp}}
\end{equation}
a theoretical model correctly describes the data if they follow the
standard normal distribution $N(0,1)$.~\footnote{More generally, if the
sufficient goodness of fit condition $\chi^2/\nu= 1 \pm \sqrt{2/\nu}$
is not fulfilled one can scale globally the errors $\Delta O_i^{\rm
  exp} \to \alpha \Delta O_i^{\rm exp} $by a common Birge factor
$\alpha$ {\it provided} the residuals follow a scaled normal
distribution $N(0,\alpha)$.
} This self-consistency condition, which can only be checked
\emph{a posteriori} and entitles legitimate error propagation, has usually been overlooked in the NN
literature. 
%A large body of normality tests is available in the
%literature and several implementations exist for different
%mathematical packages~\cite{evans2009probability,
%  james2006statistical}
%Still, \emph{any} classical statistical error analysis
%coming from a least-squares fit to data must fulfill this condition of
%normality, and only then error propagation can be confidently
%pursued. 
In~\cite{Perez:2014yla} a few of these tests
are reviewed along with a recently proposed Tail-Sensitive
test~\cite{Aldor:2013} and it was found that the three potentials
DS-OPE~\cite{Perez:2013jpa}, DS-$\chi$TPE~\cite{Perez:2013oba} and
Gauss-OPE~\cite{Perez:2014yla} have standard normal residuals. The
three more recent potentials DS-Born, Gauss-$\chi$TPE and Gauss-Born
also were found to have normally distributed
residuals~\cite{Perez:2014waa}. A simple and straightforward recipe to
apply the Tail-Sensitive test to any set of empirical data with a
sample size up to $N=9000$ was developed
in~\cite{Perez:2014kpa}. Thus, this six new potentials are the first
to qualify for error estimation in nuclear physics. A direct
application of normality is the re-sampling of experimental data via
Monte Carlo techniques, as is noted in~\cite{Perez:2014jsa}, for a
robust analysis of possible asymmetries on the potential parameters
distribution. Most recently a Monte Carlo method was used to calculate
a realistic statistical uncertainty of the Triton binding energy
stemming from NN scattering data~\cite{Perez:2014laa}.

\section{Conclusions}

In this contribution we outline the two main requirements for a
correct quantification of the NN statistical uncertainties; a correct
description of the experimental data and the reproduction of
experimental errors. Although the first one may seem obvious, some
widely used interactions in nuclear structure and nuclear reaction
calculations are fitted to phase-shifts instead of experimental data
and as shown in~\cite{Perez:2014bua} those two descriptions are not
entirely equivalent. The second requirement can be easily reformulated
into positively testing for the normality of residuals.  Normality of
residuals provides a criterion of ``falsibility'' to distinguish those
NN interactions that can come in conflict with observation and those
that cannot.  Interactions that fail the
criterion~\cite{Ekstrom:2013kea, Carlsson:2015vda} come in conflict
with observation, even if they apparently give a good description of
the data, because the chi-squared distribution cannot be used to test
goodness of fit, and require more complex statistical techniques to
analyze the data. Even though the description here is for statistical
uncertainties, a certain glance into the systematic uncertainty can
already be perceived by looking into the differences in phase-shifts
and scattering amplitude predictions given by the different realistic
potentials with $\chi^2_{\rm min}/{\rm d.o.f.}  \sim 1$. In particular
the six interactions fitted to the Granada self-consistent database,
give similar statistical uncertainties but present inconsistent
phase-shifts at low angular momentum and high energy. The
discrepancies between different potentials, accounting for the
systematic uncertainty, are usually an order of magnitude larger than
the statistical error bars.

\begin{acknowledgement}
This work is supported by Spanish DGI (grant FIS2014-59386-P) and
Junta de Andaluc{\'{\i}a} (grant FQM225). Work performed under the
auspices of the U.S. Department of Energy by Lawrence Livermore
National Laboratory under Contract No. DE-AC52-07NA27344.  Funding was
also provided by the U.S.  Department of Energy, Office of Science,
Office of Nuclear Physics under Award No.  DE-SC0008511 (NUCLEI SciDAC
Collaboration)
\end{acknowledgement}

%
% BibTeX or Biber users please use (the style is already called in the class, ensure that the "woc.bst" style is in your local directory)
%  \bibliography{references}

\begin{thebibliography}{29}

\bibitem{Ireland:0954}
D.G. Ireland, W.~Nazarewicz, J. Phys. \textbf{G42}, 030301 (2015)

\bibitem{Dobaczewski:2014jga}
J.~Dobaczewski, W.~Nazarewicz, P.G. Reinhard, J. Phys. \textbf{G41}, 074001
  (2014), \texttt{1402.4657}

\bibitem{McDonnell:2015sja}
J.D. McDonnell, N.~Schunck, D.~Higdon, J.~Sarich, S.M. Wild, W.~Nazarewicz,
  Phys. Rev. Lett. \textbf{114}, 122501 (2015), \texttt{1501.03572}

\bibitem{Furnstahl:2014xsa}
R.J. Furnstahl, D.R. Phillips, S.~Wesolowski, J. Phys. \textbf{G42}, 034028
  (2015), \texttt{1407.0657}

\bibitem{Carlsson:2015vda}
B.D. Carlsson, A.~Ekström, C.~Forssén, D.F. Strömberg, O.~Lilja, M.~Lindby,
  B.A. Mattsson, K.A. Wendt (2015), \texttt{1506.02466}

\bibitem{Perez:2014bua}
R.~Navarro~P\'erez, J.E. Amaro, E.R. Arriola, Phys. Rev. \textbf{C91}, 054002
  (2015), \texttt{1411.1212}

\bibitem{Wilson:1973jj}
K.~Wilson, J.B. Kogut, Phys.Rept. \textbf{12}, 75 (1974)

\bibitem{Bogner:2003wn}
S.~Bogner, T.~Kuo, A.~Schwenk, Phys.Rept. \textbf{386}, 1 (2003),
  \texttt{nucl-th/0305035}

\bibitem{Bogner:2009bt}
S.~Bogner, R.~Furnstahl, A.~Schwenk, Prog.Part.Nucl.Phys. \textbf{65}, 94
  (2010), \texttt{0912.3688}

\bibitem{Aviles:1973ee}
J.~Aviles, Phys.Rev. \textbf{C6}, 1467 (1972)

\bibitem{NavarroPerez:2011fm}
R.~Navarro~P\'erez, J.E. Amaro, E.~Ruiz~Arriola, Prog.Part.Nucl.Phys.
  \textbf{67}, 359 (2012), \texttt{1111.4328}

\bibitem{Gezerlis:2014zia}
A.~Gezerlis, I.~Tews, E.~Epelbaum, M.~Freunek, S.~Gandolfi et~al. (2014),
  \texttt{1406.0454}

\bibitem{Lynn:2014zia}
J.E. Lynn, J.~Carlson, E.~Epelbaum, S.~Gandolfi, A.~Gezerlis, A.~Schwenk, Phys.
  Rev. Lett. \textbf{113}, 192501 (2014), \texttt{1406.2787}

\bibitem{Gandolfi:2014ewa}
S.~Gandolfi, A.~Lovato, J.~Carlson, K.E. Schmidt, Phys. Rev. \textbf{C90},
  061306 (2014), \texttt{1406.3388}

\bibitem{Stoks:1993tb}
V.~Stoks, R.~Kompl, M.~Rentmeester, J.~de~Swart, Phys.Rev. \textbf{C48}, 792
  (1993)

\bibitem{Gross:2008ps}
F.~Gross, A.~Stadler, Phys.Rev. \textbf{C78}, 014005 (2008), \texttt{0802.1552}

\bibitem{Perez:2013jpa}
R.~Navarro~P\'erez, J.E. Amaro, E.~Ruiz~Arriola, Phys.Rev. \textbf{C88}, 064002
  (2013), \texttt{1310.2536}

\bibitem{GranadaDB}
R.~Navarro~Perez, J.~Amaro, E.~Ruiz~Arriola, \emph{{2013 Granada Database}},
  \url{http://www.ugr.es/~amaro/nndatabase/} (2013), accessed: 2015-08-15

\bibitem{Perez:2013mwa}
R.~Navarro~P\'erez, J.E. Amaro, E.~Ruiz~Arriola, Phys.Rev. \textbf{C88}, 024002
  (2013), \texttt{1304.0895}

\bibitem{evans2009probability}
M.~Evans, J.~Rosenthal, \emph{Probability and Statistics: The Science of
  Uncertainty} (W. H. Freeman, 2009), ISBN 9781429224628

\bibitem{james2006statistical}
F.~James, W.~Eadie, \emph{Statistical Methods in Experimental Physics} (World
  Scientific, 2006), ISBN 9789812567956

\bibitem{Perez:2014yla}
R.~Navarro~P\'erez, J.E. Amaro, E.~Ruiz~Arriola, Phys.Rev. \textbf{C89}, 064006
  (2014), \texttt{1404.0314}

\bibitem{Aldor:2013}
S.~Aldor-Noiman, L.D. Brown, A.~Buja, W.~Rolke, R.A. Stine, Am. Statist.
  \textbf{67}, 249 (2013)

\bibitem{Perez:2013oba}
R.~Navarro~P\'erez, J.E. Amaro, E.~Ruiz~Arriola, Phys.Rev. \textbf{C89}, 024004
  (2014), \texttt{1310.6972}

\bibitem{Perez:2014waa}
R.~Navarro~P\'erez, J.~Amaro, E.R. Arriola (2014), \texttt{1410.8097}

\bibitem{Perez:2014kpa}
R.~Navarro~P\'erez, J.E. Amaro, E.~Ruiz~Arriola, J. Phys. \textbf{G42}, 034013
  (2015), \texttt{1406.0625}

\bibitem{Perez:2014jsa}
R.~Navarro~P\'erez, J.E. Amaro, E.~Ruiz~Arriola, Phys.Lett. \textbf{B738}, 155
  (2014), \texttt{1407.3937}

\bibitem{Perez:2014laa}
R.~Navarro~P\'erez, E.~Garrido, J.~Amaro, E.~Ruiz~Arriola, Phys.Rev.
  \textbf{C90}, 047001 (2014)

\bibitem{Ekstrom:2013kea}
A.~Ekstr{\"o}m, G.~Baardsen, C.~Forss{\'e}n, G.~Hagen, M.~Hjorth-Jensen et~al.,
  Phys.Rev.Lett. \textbf{110}, 192502 (2013), \texttt{1303.4674}

\end{thebibliography}
%
% Non-BibTeX users please use
%

\end{document}